# Stellar mass loss-driven wind models of elliptical galaxies?

B.K. Gibson[1,2]
[1] *Department of Astrophysics, University of Oxford, Keble Road, Oxford, UK OX1 3RH*
[2] *Department of Geophysics & Astronomy, University of British Columbia, Vancouver, British Columbia, Canada V6T 1Z4*

**ABSTRACT**
Recent claims in the literature (Bressan, Chiosi & Fagotto 1994) that the epoch describing the onset of galactic winds in spheroidal star systems has been severely overestimated in the past, due to the neglect of energy deposited in the interstellar medium from stellar winds, is evaluated in the light of a more conservative approach to modeling the input of thermal energy to the system from massive stars undergoing mass loss. Applying the most recent models of stellar kinetic energy deposition, coupled with reasonable assumptions regarding the efficiency of thermalisation, it is shown that contrary to the aforementioned study, the influence of stellar winds in driving the galactic winds in ellipticals of $M_{\rm G} \gtrsim 10^9$ $M_\odot$ is most likely negligible.

## 1 INTRODUCTION

Within the well-established analytical framework (e.g. Matteucci & Tornambé 1987 and Arimoto & Yoshii 1987) for studying the spectral and chemical evolution of spheroidal systems, and their dependence upon the coupled evolution of the interstellar gas thermal energy, the recent work of Bressan, Chiosi & Fagotto (1994) (hereafter BCF94) is perhaps the most comprehensive to date.

This framework posits that active star formation occurs during the early stages of elliptical evolution, leading to a high rate of supernovae (SNe) which triggers galactic winds once the residual thermal energy of all SNe remnants reaches the binding energy of the system's gas. This remaining gas is expelled and star formation halted, the subsequent evolution being regulated by the gas returned to the interstellar medium (ISM) by "dying" stars. Such supernova-driven wind models were originally postulated by Mathews & Baker (1971), and successfully exploited in explaining the observed correlations among properties of spheroidal systems (e.g. Larson 1974, Faber 1977, Saito 1979), including the mass-metallicity relationship.

Predicting the epoch of galactic wind onset $t_{\rm GW}$ is crucial (Angeletti & Giannone 1990), as $t_{\rm GW}$ governs the stop of star formation, mass of expelled metals and gas, and current photometric properties of ellipticals. Two recognised key parameters for estimating $t_{\rm GW}$ are: (1) the initial protogalactic radius (which sets the binding energy of the system), and (2) the evolution of the thermal energy content of the SNe remnants. The former has been addressed by Angeletti & Giannone (1991) and the latter by Gibson (1994), and are not wholly relevant for the discussion which follows as we will simply adopt the formalism used by BCF94.

One extraordinary claim of BCF94 is that the dominant mechanism for setting $t_{\rm GW}$ in these analytical models is the thermal energy deposited into the ISM by the stellar winds associated with mass loss from massive stars. In their work, this component is of such a magnitude that SNe energy becomes a negligible contributor, and $t_{\rm GW}$ occurs extremely early during a galaxy's lifetime. e.g. for a $10^{12}$ $M_\odot$ elliptical, BCF94 predict $t_{\rm GW}$=0.09 Gyr, as opposed to others who find $t_{\rm GW} \sim 1.5$ Gyr (e.g. Angeletti & Giannone 1990 and Padovani & Matteucci 1993), although early wind epochs are also favoured by Elbaz, Arnaud & Vangioni-Flam (1994). In their study, though, the responsible factor is an adopted initial mass function which is truncated at a high lower mass limit ($m_{\rm L} = 3$ $M_\odot$), as opposed to stellar wind energy.

While it has been recognised that stellar wind energy deposition may play a role in setting the timescale for starbursts of mass $\lesssim 10^7$ $M_\odot$ (e.g. Leitherer, Robert & Drissen 1992), the BCF94 results are the first to postulate that they regulate the epoch of gas ejection for massive ellipticals up to $\sim 3 \times 10^{12}$ $M_\odot$. We feel that such a remarkable conclusion merits further study, and the short analysis presented herein is meant to be a first tentative step in that direction.

Our aim is not to take strong issue with the BCF94 conclusion, but only to draw attention to the assumptions in their stellar wind energy formalism with an eye towards understanding the source of their very small galactic wind times, and present an alternative, more conservative model, in order to illustrate that *perhaps* the importance of the contribution of stellar winds in driving the onset of galactic winds has been overestimated in their study. Section 2 briefly outlines the chemical evolution and basic wind mod-els used in our work, as well as the methodology adopted to incorporate stellar wind energy deposition. Section 3 includes a discussion of our conclusions and their relevance in light of the BCF94 results.

## 2  ANALYSIS

### 2.1  The classical wind model of ellipticals

Following BCF94, we employ the classical one-zone wind model, introduced in its modern form by Matteucci & Greggio (1986), as the framework in which to monitor the evolution of the gas mass, metallicity, and thermal energy of spheroidal systems. The fundamental equations, assumptions, and input ingredients relevant to the specific package used here (entitled **MEGaW**≡**M**etallicity **E**volution with **Ga**lactic **W**inds) are described in detail by Gibson (1995), although much of the relevant background material can also be found in Matteucci & Tornambé (1987), Arimoto & Yoshii (1987), or Angeletti & Giannone (1990).

The star formation rate $\psi(t)$, as in BCF94, is taken to be directly proportional to the available gas mass $M_g(t)$, i.e.

$$\psi(t) = \nu M_g(t) \quad [M_\odot/\text{yr}], \tag{1}$$

where the constant of proportionality $\nu$ represents the inverse of the star formation timescale. Typically one assumes this measure of the star formation efficiency is a function of the galactic mass $M_G$:

$$\nu = 8.6 (M_G/10^{12} M_\odot)^{-0.115} \quad [\text{Gyr}^{-1}], \tag{2}$$

which stems from assuming the initial timescale for star formation is set by the mean collision time of star forming fragments in the proto-galaxy (Arimoto & Yoshii 1987). BCF94 neglect the mass dependency inherent in equation 2 and simply treat $\nu$ as a free parameter. Unless explicitly stated that equation 2 has been used, we shall adopt their preferred model which claims a universal value of $\nu = 20.0$ Gyr$^{-1}$.

Again following BCF94, we take the initial mass function (IMF) by mass $\phi(m)$ to be the singular power law of slope $x = 1.35$ (Salpeter 1955), normalised to unity with an upper mass limit of 120.0 $M_\odot$, and a lower limit $m_L$ selected such that the fraction of the IMF stored above 1.0 $M_\odot$, is 2/5. i.e. $m_L = 0.1$ $M_\odot$.

For the work presented here we have used the yields of Arnett (1991) for Type II SNe, Renzini & Voli (1981) for single low and intermediate mass stars, and Thielemann, Nomoto & Yokoi (1986) for Type Ia SNe. Details of the prescriptions used and the assumed SNe progenitors can be found in Gibson (1995).

A direct comparison of metallicity predictions from our package and that of BCF94 is difficult, as they have not listed the source of their yields, nor do they incorporate the input from Type Ia SNe, a problem which they themselves recognise in their Section 7.4. As such, a detailed analysis of the differences between the metal production in the two codes will not be attempted here, nor is it necessary. By simply demonstrating the sensitivity of $t_{\rm GW}$ to the stellar wind energy formalism, one can estimate qualitatively the effect upon the mass of metals ejected during the galactic wind phase, as well as the metal content of the remaining stellar population, by recognising that these abundances are proportional to $t_{\rm GW}$.

Where appropriate, we have used the analytical fits for main sequence lifetimes given by Güsten & Mezger (1982), to set the time at which SN energy for a star of mass $m$ is injected.

In the classical wind model, for gas to be expelled from a galaxy, the thermal energy of the gas $E_{\rm th_{SN}}$, heated by SNe explosions, should exceed the binding energy of the gas $\Omega_{\rm gas}$ (Larson 1974). i.e. the galactic wind will start at the time $t_{\rm GW}$ when:

$$E_{\rm th_{SN}}(t_{\rm GW}) = \Omega_{\rm gas}(t_{\rm GW}). \tag{3}$$

In one-zone models, $\Omega_{\rm gas}(t)$ can be influenced by assumptions regarding the distribution of dark matter, although for realistic diffuse halo distributions Matteucci (1992) finds that the effect upon $t_{\rm GW}$ is negligible. Hence we simply use the expression for the evolution of $\Omega_{\rm gas}$ from Arimoto & Yoshii (1987) and Saito (1979).

The total thermal energy in the gas at time $t$ is given by

$$E_{\rm th_{SN}}(t) = \int_0^t \varepsilon_{\rm th_{SN}}(t - t') R_{\rm SN}(t') dt', \tag{4}$$

where $t'$ is the explosion time, $R_{\rm SN}(t) = R_{\rm SNIa}(t) + R_{\rm SNII}(t)$ is the sum of the Type Ia and II SNe rates, and $\varepsilon_{\rm th_{SN}}$ is the equation governing the evolution of the thermal energy content in the interior of a SN remnant, taken directly from Cox (1972) and Chevalier (1974). The Type Ia and II SNe rate formulae are from Greggio & Renzini (1983) and Matteucci & Tornambé (1987).

### 2.2  Energy deposition from stellar winds

By analogy with the formalism used to calculate the residual SNe thermal energy as a function of time, we can write the thermalised kinetic energy in the ISM $E_{\rm th_W}$ due to mass loss from massive ($m \gtrsim 12$ $M_\odot$) stars as

$$E_{th_W}(t) = \eta \int_0^t \int_{12.0}^{m_U} \frac{\phi(m)}{m} \psi(t') \varepsilon_W\left(m, t - t', Z(t')\right) dm dt'. \tag{5}$$

$\varepsilon_W$ is the kinetic energy deposited by a star of mass $m$ and metallicity $Z$ as a function of time $t - t'$ since its zero age main sequence. As we will show, knowledge of $\varepsilon_W$ and its evolution can be a primary factor in setting the time for global gas ejection.

Another key ingredient in predicting $t_{\rm GW}$ is the assumed kinetic energy thermalisation efficiency parameter $\eta$. The evolution of adiabatic interstellar bubbles due to stellar winds from high mass stars interacting with the interstellar medium, including radiative losses, imply thermalisation efficiencies in the range $\eta = 0.2 \rightarrow 0.4$ (e.g. Weaver et al. 1977 and Koo & McKee 1992), with $\eta \sim 0.3$ being a fair compromise (Leitherer 1994). Contrary to this, BCF94 have set $\eta = 1$ (i.e. all the stellar wind kinetic energy is thermalised). The sensitivity of $t_{\rm GW}$ to this assumed parameter $\eta$ is discussed in Section 3.

As equation 5 intimates, we must assume some form for the evolution of a star's kinetic energy output to the ISM $\varepsilon_W$ as a function of time. To generate such an appropriate grid of models, we use the recent results of Leitherer, Robert &

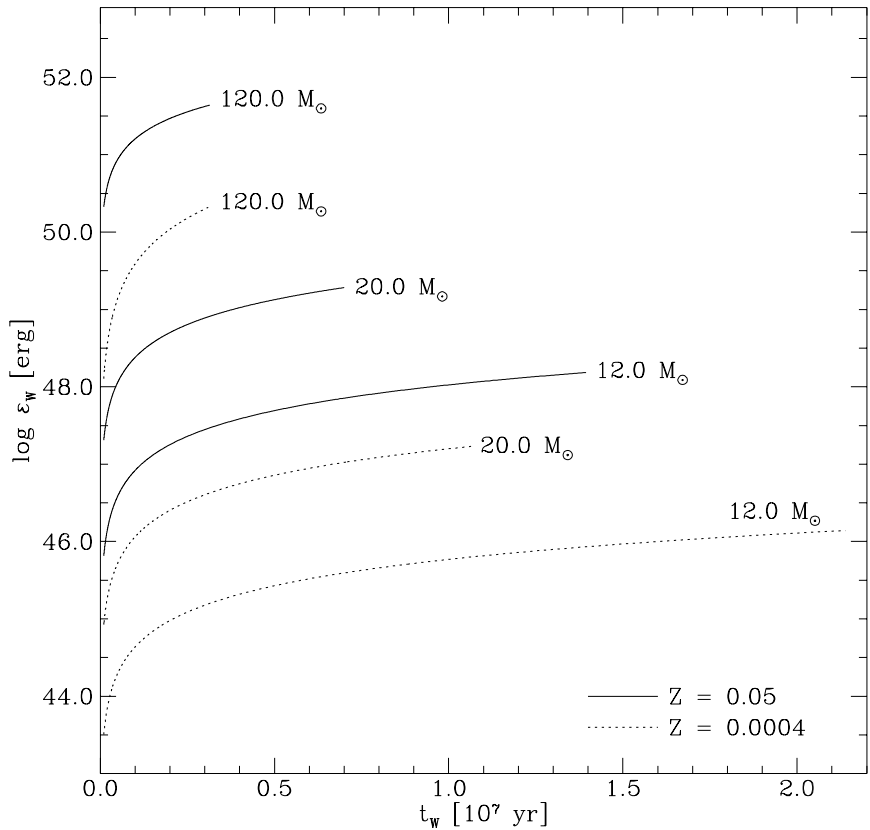

**Figure 1.** Total kinetic energy deposited as a function of time from the zero age main sequence to the end of the star's lifetime for the extrema in mass and metallicity of interest in this work. A fraction $\eta$ of the wind's kinetic energy is thermalised by development of shocks in the zone where winds interact with the surrounding interstellar medium.

Drissen (1992) (hereafter LRD92), who compute a series of radiative wind models of hot stars and derive various scaling relations for the mass loss rates $\dot{m}$ and wind velocities $v_\infty$ as functions of stellar mass, metallicity, luminosity, and effective temperature, for all important stages of post-main sequence evolution, including the OB, luminous blue variable (LBV), red supergiant, and Wolf-Rayet (W-R) phases. The LRD92 Mass-Loss Model C provides the necessary relations for $\dot{m}$ and $v_\infty$ during these four phases. LRD92 then couple these relations to the Maeder (1990) grid of stellar evolution tracks to assign appropriate timescales for each phase. We have chosen to use the more extensive modern suite of evolution tracks provided by the Padova Group (Fagotto et al. 1994), and have adjusted modestly the effective temperature discriminator for the LBV/W-R phases to better represent this newer grid of tracks. Specifically, LRD92 assume the LBV/W-R phase transition in the (Maeder 1990) tracks is best represented by $\log T_{\text{eff}} \sim 4.4$, whereas in the Padova tracks, $\log T_{\text{eff}} \sim 4.0$ seems more appropriate. In addition, as opposed to breaking down the W-R phase into five sub-classes as done in LRD92, we have simply chosen two, leaving the WN classification to stand alone, but combining the WC and WO sub-classes. Both modifications were chosen for the sake of simplicity and do not alter our results or conclusions.

A grid of 35 $\varepsilon_W$ versus $t$ curves were generated (five metallicities from $Z = 0.0004 \rightarrow 0.05$ and seven masses from $m = 12.0 \rightarrow 120.0$ M$_\odot$) using the procedure outlined above, a small sample of which are shown in Figure 1.

One final point which must be addressed is the fate of the stellar wind thermal energy after the end of the star's lifetime $\eta\varepsilon_W(t > \tau_m)$ (i.e. the point at which the SN energy injection commences). This is somewhat more problematic as, unlike that done for SNe remnant thermal energy in the post-shell formation regime (Cox 1972), there has yet to be a similar study done for massive stars in the post-SNe phase. We shall adopt a conservative model (Heckman 1994 and Leitherer 1994) in which the residual thermal energy within the stellar wind-induced interstellar bubble at the time of SN explosion $\eta\varepsilon_W(t \equiv \tau_m)$ is added to the thermal energy initially supplied by the SN ($\varepsilon_{\text{th}_{\text{SN}}}(t_{\text{SN}} \equiv 0) \approx 7 \times 10^{50}$ erg), and thereafter the remnant taken to evolve like the typical Cox (1972) SN remnant, but with this additional component of thermal energy (hereafter referred to as Model D). To first order, such an assumption is supported by the studies of Tenorio-Tagle et al. (1990) governing the evolution of SN remnant thermal energy within pre-existing stellar wind-driven bubbles in the ISM (e.g. their Figures 1 and 11). Note that this added component will be a function of the progenitor star's mass and metallicity, as witnessed by the end points for the various models shown in Figure 1.

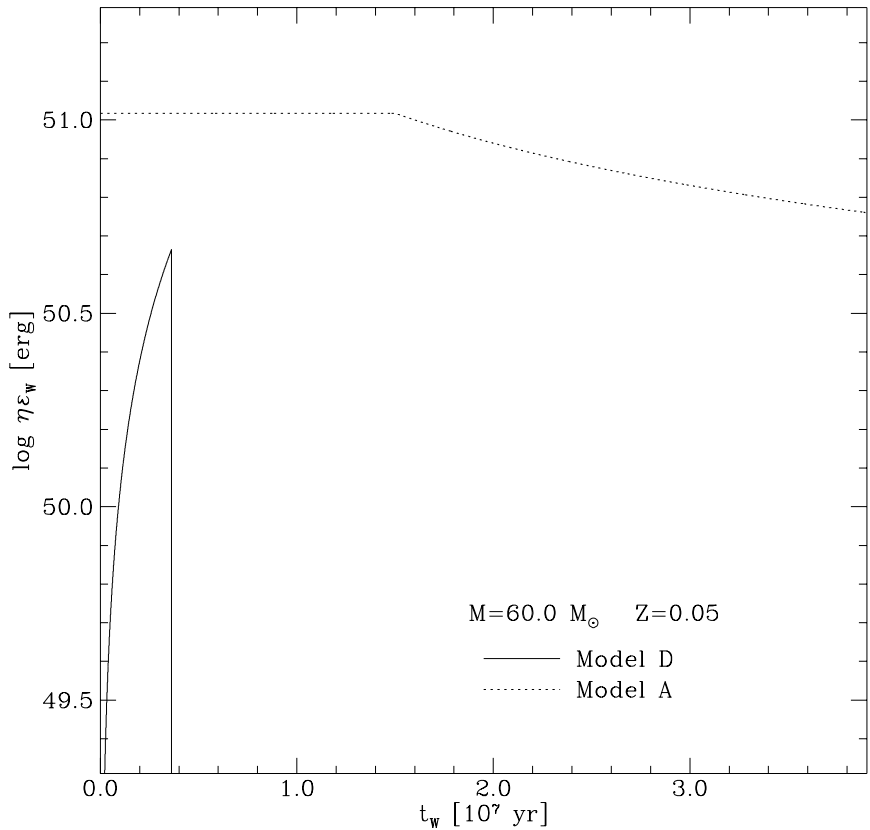

**Figure 2.** Thermal energy available in the interstellar medium as a function of time due to a $m = 60.0$ M$_\odot$, $Z = 0.05$ star undergoing mass loss via stellar winds. The dotted line represents the formalism used by BCF94 (Model A) whereas the solid line denotes the most conservative model (Model D) described in our study.

Such a model is consistent with preliminary work by Matteucci (1994) suggesting that the stellar wind kinetic energy thermalisation efficiency approaches zero within the first few million years of energy deposition. Model C is identical to Model D except that the astration parameter, as in BCF94, is simply set to $\nu = 20.0$ Gyr$^{-1}$, as opposed to the appropriate value determined from equation 2. Model E represents the results of Model D, but with the contribution of stellar winds ignored. i.e. $\eta \varepsilon_W(t) = 0.0$, for all times.

To compare our model predictions against those of BCF94, we have replicated their formalism and denoted it as Model A. By analogy with the behaviour of a SN remnant's thermal energy evolution, which decreases as $\varepsilon_{th_{SN}} \propto (t/t_c)^{-0.62}$ for times longer than the cooling time $t_c$ (Cox 1972), Model A assumes $\varepsilon_W \propto (t/t_{cw})^{-0.62}$ for $t > t_{cw}$, where $t_{cw}$ is taken simply to be the universal, and somewhat arbitrary, value $1.5 \times 10^7$ yr, as opposed to the more appropriate choice of $t_{cw} \equiv \tau_m$. For $t \leq t_{cw}$, BCF94 assume

$$\varepsilon_W = 1.989 \times 10^{49} m \left(\frac{Z}{0.06}\right)^{3/4} \quad [\text{erg}], \qquad (6)$$

where $m$ is the initial stellar mass in solar masses. Note that unlike the SN post-shell formation thermal energy which initiates at $t = t_c$ at a level which is $\sim 30\%$ (Cox 1972) that of the initial SN thermal energy and decreases following the power law described above, BCF94 take the initial residual stellar wind energy at $t = t_{cw}$ to be 100% that given by equation 6. It is also important to recall that while our models assume a thermalisation efficiency $\eta = 0.3$, BCF94 have taken $\eta = 1.0$, contrary to the evidence presented earlier. Model B is equivalent to Model A, except neglecting the contribution of residual SNe thermal energy - i.e. only stellar wind energy included. The primary models, and their respective parameters, are listed in Table 1.

In summary, an important distinction between our models and those of BCF94 comes in the definition of the aforementioned stellar wind "cooling time". Whereas we have properly set the timescale of energy deposition to be equal to a star's lifetime, and accounted for the differing mass loss rates and ejection velocities according to a given star's post-main sequence evolutionary status, BCF94 have simply set this timescale, regardless of mass or metallicity, to be a constant $t_{cw} = 1.5 \times 10^7$ yrs, and assumed that the thermal energy is constant during this time and given by equation 6. Coupled with this, BCF94 have set the kinetic energy thermalisation efficiency parameter $\eta$ to unity, contrary to Weaver et al. (1977) who show $\eta \approx 0.3$, a point corroborated by Koo & McKee (1992) and Leitherer (1994). The obvious differences in our approaches can be seen quite graphically in Figure 2. A 60.0 M$_\odot$, $Z = 0.05$ star's residual thermal energy contribution following our formalism (solid line) is shown in relation to that used by BCF94 (dotted line). Here we can see that both the magnitude, shape, and duration of their stellar wind thermal energy contribution is grossly overestimated, in both the pre-SN and post-SN regimes, relative to our models.

**Table 1.** The input ingredients for the primary models discussed in this paper. Model A represents the formalism used by Bressan, Chiosi & Fagotto (1994), whereas Model D denotes our preferred conservative model. Note: GM82 = Güsten & Mezger (1982) and C72+$\eta\varepsilon_W(\tau_m)$ = Cox (1972) plus an additional component of residual stellar wind thermal energy set by the end points of the appropriate curves in Figure 1.

| Model | $t_{cw}$ Gyr | $\eta$ | $\nu$ Gyr$^{-1}$ | $\varepsilon_W$ $t \leq t_{cw}$ | $\eta\varepsilon_W$ $t > t_{cw}$ | $\varepsilon_{th_{SN}}$ |
|---|---|---|---|---|---|---|
| A | 0.015 | 1.0 | 20.0 | eqn 6 | fig 2 | C72 |
| B | 0.015 | 1.0 | 20.0 | eqn 6 | fig 2 | 0.0 |
| C | GM82 | 0.3 | 20.0 | fig 1 | 0.0 | C72+$\eta\varepsilon_W(\tau_m)$ |
| D | GM82 | 0.3 | eqn 2 | fig 1 | 0.0 | C72+$\eta\varepsilon_W(\tau_m)$ |
| E | GM82 | 0.0 | eqn 2 | 0.0 | 0.0 | C72 |

We mention in passing that other sources of thermal energy input have not been considered in past models of this sort, nor are they done so here. Neither mass loss during the thermally pulsing regime of the asymptotic giant branch for low mass stars (Vassiliadis & Wood 1993), nor envelope ejection in the planetary nebulae phase (Van Buren 1985) are important contributors to the system's thermal energy. This has been attributed to the extremely small ejection velocities involved in both cases ($\lesssim 50$ km/s), compared for example with those from stellar winds ($\sim 2000$ km/s).

## 3 DISCUSSION

In Table 2 we present estimates for the time of galactic wind onset $t_{GW}$ as a function of initial galaxy mass for the models described in Section 2. Recall that Model A represents the expected results when a formalism for the residual stellar wind thermal energy $\eta\varepsilon_W(t)$ as used by BCF94 is assumed. The results as given by BCF94 are actually $\sim 20\%$ smaller than these, and identifying the source of this difference is difficult, but not particularly important to the arguments presented here.

Recalling the severe differences in formalism for $\varepsilon_W(t)$ as used by BCF94 (Model A) versus the conservative Model D of Section 2, as evidenced by Figure 2, it is not surprising to see that $t_{GW}$ occurs substantially earlier in the BCF94 analysis. e.g. 0.115 Gyr versus 1.603 Gyr for the $10^{12}$ M$_\odot$ case. Of course, part of the difference between the Model A and D numbers is simply due to the assumed value of the astration parameter $\nu$. If instead of using equation 2, we simply set $\nu = 20.0$ Gyr$^{-1}$ as was done by BCF94, we can see from the results for Model C that $t_{GW}$ is still four to eight times greater for all galactic masses, compared with those of BCF94.

An interesting result is seen if we look at the BCF94 (Model A) predictions but in the absence of any thermal energy contribution from SNe. This is what is denoted by Model B. Comparing Model B with the BCF94 predictions for $t_{GW}$, we see that there is virtually no difference. In fact, we can conclude that in the BCF94 formalism, the sole driving mechanism for the onset of galactic winds is residual stellar wind thermal energy. SNe are an entirely negligible component, contrary to what has been reported in all previous work (e.g. Larson 1974, Arimoto & Yoshii 1987, Matteucci & Tornambé 1987, Angeletti & Giannone 1990). If we neglect SNe in our Model D, galactic winds are never induced, except in the $10^7$ M$_\odot$ case. As expected (e.g. Leitherer, Robert & Drissen 1992), we find that stellar winds can regulate the onset of global winds at this mass scale. For $m \gtrsim 10^9$ M$_\odot$, $t_{GW}$, in our models, is not strongly influenced by stellar winds. This is corroborated by comparing Model D predictions with those of Model E, which as we recall from Section 2, simply represents the Model D predictions without any wind contribution. A similar conclusion was found for Model C predictions in the absence of any wind component.

**Table 2.** A comparison of galactic wind epochs $t_{GW}$ (in units of Gyr) for the models described in the text. Model A represents the predictions of Bressan et al. 1994 (BCF94) for initial masses ranging from $10^7 \to 10^{12}$ M$_\odot$, while Model D is our most conservative model.

| Model | $1.0 \times 10^7$ | $1.0 \times 10^9$ | $5.0 \times 10^{10}$ | $1.0 \times 10^{12}$ |
|---|---|---|---|---|
| A | 0.011 | 0.022 | 0.051 | 0.115 |
| B | 0.011 | 0.023 | 0.056 | 0.123 |
| C | 0.041 | 0.159 | 0.279 | 0.451 |
| D | 0.022 | 0.160 | 0.566 | 1.603 |
| E | 0.032 | 0.181 | 0.566 | 1.603 |

We can identify three contributing factors for the extremely short timescales for global wind ejection as presented in the BCF94 paper, each of which are self-evident upon re-examination of Figure 2. First, BCF94 have assumed that all the energy injected by stellar winds into the ISM is in the form of thermal energy, whereas we have already referred to the fact that a value of $\sim 30\%$ seems more appropriate. Second, BCF94 assume that all massive stars, regardless of mass or metallicity, have made available some constant amount of thermal energy, governed by equation 6, for $1.5 \times 10^7$ yrs. This is wrong on several accounts: (i) this timescale is substantially longer than the lifetime of every massive star in their computations. e.g. the 60.0 M$_\odot$ star of Figure 2 has a lifetime a factor of five smaller than the universal value used by BCF94; (ii) BCF94 have neglected the fact that the energy contribution during the lifetime is not constant and is indeed a function of the post-main sequence evolutionary status. Third, BCF94 have assumed a very slow decline in the residual stellar wind thermal energy during the post-SN phase, whereas the evidence presented in Section 2 seems to suggest that this is not true, and in fact, this residual energy can be reasonably approximated by setting it to zero during the post-SN phase.

On a more general note, it is important to stress that the homogeneous single zone ISM assumption inherent to most of the galactic wind codes discussed earlier, including

BCF94, is at best a crude representation of reality. It is apparent that ellipticals show radial gradients in many of their observable properties (e.g. mass density, metallicity, escape velocity), each of which impacts upon the ISM thermal energy evolution. For example, the radiative cooling time of an individual remnant's shell, as well as its hot dilute interior, depends upon both chemical composition and local mass density (e.g. Franco et al. 1994 and Gibson 1994). A recent attempt to take into account the influence of a de Vaucouleurs gas density profile in computing the global ISM energy can be found in Elbaz, Arnaud & Vangioni-Flam (1994). Adopting a multi-phase ISM (e.g. cold cloud nuclei, surrounded by warm neutral gas, embedded in a hot thin corona – Ostriker & McKee 1988) would also be a further step in the right direction but in practice becomes increasingly difficult to do so within the framework of the classic analytical wind model without the introduction of a greater number of free parameters. In this vein, the work of Ferrini & Poggianti (1993) is a very important first attempt at modelling such multi-phase ISMs while retaining the simple elegance of the analytical galactic wind framework. Similarly, hydrodynamical simulations of galactic winds driven through "cloudy" (e.g. Missoulis 1994) and/or "smooth" (e.g. David, Forman & Jones 1990, Ciotti et al. 1991) ISMs in ellipticals must continue to be a primary aspect of future work.

Finally, even within the context of the simple one-zone model of BCF94, we do not mean to suggest that we have explored fully the parameter space of wind solutions – e.g. the influence of low mass IMF truncation (Elbaz, Arnaud & Vangioni-Flam 1994); the adoption of a flatter (e.g. Arimoto & Yoshii 1987) or steeper (e.g. Scalo 1986) IMF; steady cosmic ray-driven galactic winds (e.g. Ipavich 1975); magnetic field energy in ellipticals (Lesch & Bender 1990). Bearing in mind the range of "free" parameters in such models (e.g. IMF slope and bounds, astration parameter, SN remnant thermal energy formalism, dark matter distribution), it would not be surprising to find that the early wind epochs preferred by BCF94 are a possibility. Indeed, taking a different approach, Elbaz, Arnaud & Vangioni-Flam (1994) found similarly small values for $t_{\rm GW}$.

Thus, in conclusion, we re-iterate that our goal in this paper has not been to argue against the early BCF94 wind epochs per se, but simply to demonstrate that adopting a more conservative and realistic formalism for the stellar wind thermal energy, leads to the conclusion that stellar winds play only a negligible role in determining the onset of galactic winds in normal and giant ellipticals.

## ACKNOWLEDGEMENTS

I would like to thank Cedric Lacey, Claus Leitherer, Tim Heckman, and Francesca Matteucci for a number of useful discussions and correspondences.

## REFERENCES

Angeletti, L., Giannone, P., 1990, A&A, 234, 53
Angeletti, L., Giannone, P., 1991, A&A, 248, 45
Arimoto, N., Yoshii, Y., 1987, A&A, 173, 23
Arnett, D., 1991, in Lambert, D.L., ed, Frontiers of Stellar Evolution, ASP, San Francisco, p. 389
Bressan, A., Chiosi, C., Fagotto, F., 1994, ApJS, 94, 63
Chevalier, R.A., 1974, ApJ, 188, 501
Ciotti, L., d'Ercole, A., Pellegrini, S., Renzini, A., 1991, ApJ, 376, 380
Cox, D.P., 1972, ApJ, 178, 159
David, L.P., Forman, W., Jones, C., 1990, ApJ, 359, 29
Elbaz, D., Arnaud, M., Vangioni-Flam, E., 1994, A&A, in press
Faber, S.M., 1977, in Tinsley, B.M. & Larson, R.B., eds, The Evolution of Galaxies and Stellar Populations, Yale Obs., New Haven, p. 157
Fagotto, F., Bressan, A., Bertelli, G., Chiosi, C., 1994, A&AS, 104, 365
Ferrini, F., Poggianti, B.M., 1993, ApJ, 410, 44
Franco, J., Miller III, W.W., Arthur, S.J., Tenorio-Tagle, G., Terlevich, R.J., 1994, ApJ, in press
Gibson, B.K., 1994, JRASC, in press
Gibson, B.K., 1995, in preparation
Greggio, L., Renzini, A., 1983, A&A, 118, 217
Güsten, R., Mezger, P.G., 1982, Vistas Astron., 26, 159
Heckman, T.M. 1994, priv comm
Ipavich, F.M., 1975, ApJ, 196, 107
Koo, B.-C., McKee, C.F., 1992, ApJ, 388, 103
Larson, R.B., 1974, MNRAS, 169, 229
Leitherer, C., 1994, priv comm
Leitherer, C., Robert, C., Drissen, L., 1992, ApJ, 401, 596
Lesch, H., Bender, R., 1990, A&A, 233, 417
Maeder, A., 1990, A&AS, 84, 139
Mathews, W.G., Baker, J.C., 1971, ApJ, 170, 241
Matteucci F., 1992, ApJ, 397, 32
Matteucci F., 1994, in preparation
Matteucci, F., Greggio, L., 1986, A&A, 154, 279
Matteucci, F., Tornambè, A., 1987, A&A, 185, 51
Missoulis, V., 1994, Astr. Rep., 38, 12
Ostriker, J.P., McKee, C.F., 1988, Rev. Mod. Phys., 60, 1
Padovani, P., Matteucci, F., 1993, ApJ, 416, 26
Renzini, A., Voli, M., 1981, A&A, 94, 175
Saito, M., 1979, PASJ, 31, 193
Salpeter, E.E., 1955, ApJ, 121, 161
Scalo, J.M., 1986, Fund. Cosm. Phys., 11, 1
Tenorio-Tagle, G., Bodenheimer, P., Franco, J., Różyczka, M., 1990, MNRAS, 244, 563
Thielemann, F.-K., Nomoto, K., Yokoi, K., 1986, A&A, 158, 17
Vassiliadis, E., Wood, P.R., 1993, ApJ, 413, 641
Van Buren, D., 1985, ApJ, 294, 567
Weaver, R., McCray, R., Castor, J., Shapiro, P., Moore, R., 1977, ApJ, 218, 377